\documentclass[12pt,letterpaper,english,preprint, aps,superscriptaddress]{revtex4}
\usepackage[T1]{fontenc}
\usepackage[latin9]{inputenc}
\usepackage[dvipdf]{color}
\usepackage{amsmath}
\usepackage{amssymb}
\usepackage{epstopdf}
\usepackage{graphicx}



\usepackage{babel}
\begin{document}

\title{Probing the Effects of Dimension-eight Operators Describing Anomalous Neutral Triple Gauge Boson Interactions at FCC-hh}

\author{A. Senol}
\email{senol_a@ ibu.edu.tr}
\affiliation{Department of Physics, Bolu Abant Izzet Baysal University, Bolu,
Turkey.}
\author{H. Denizli}
\email{denizli_h@ ibu.edu.tr}
\affiliation{Department of Physics, Bolu Abant Izzet Baysal University, Bolu,
Turkey.}

\author{A. Yilmaz}
\email{aliyilmaz@giresun.edu.tr}

\affiliation{Department of Electrical and Electronics Engineering, Giresun University, 28200 Giresun, Turkey}

\author{I. Turk Cakir}
\email{ilkay.turk.cakir@cern.ch}
\affiliation{Department of Energy Systems Engineering, Giresun University, 28200 Giresun, Turkey}

\author{K.Y. Oyulmaz}
\email{kaan.oyulmaz@gmail.com}
\affiliation{Department of Physics, Bolu Abant Izzet Baysal University, Bolu,
Turkey.}

\author{O. Karadeniz}
\email{ozgunkdz@gmail.com}
\affiliation{Department of Physics, Bolu Abant Izzet Baysal University, Bolu,
Turkey.}
\author{O. Cakir}
\email{ocakir@science.ankara.edu.tr}
\affiliation{Department of Physics, Ankara University, 06100, Ankara, Turkey.}
\date{\today}\date{\today}
\begin{abstract}
The effects of dimension-eight operators giving rise to anomalous neutral triple gauge boson interactions of $Z\gamma\gamma$ and $Z\gamma Z$ vertices in $pp\to l^-l^+\gamma$ and $pp\to \nu\bar \nu \gamma$ are investigated at 100 TeV centre of mass energy of future circular hadron collider (FCC-hh). The  transverse momentum of photon, invariant mass of $l^-l^+\gamma$ and angular distribution of charged lepton in the rest frame of $l^-l^+$ and Missing Energy Transverse (MET) are considered in the analysis. The realistic detector effects are also included with Delphes simulation. Sensitivity limits obtained at 95\% C.L. for $C_{\widetilde B W}/\Lambda^4$ and $C_{B B}/\Lambda^4$ couplings are $[-0.52;0.52] ([-0.40;0.40])$ TeV$^{-4}$, $[-0.43;0.43] ([-0.33;0.33])$ TeV$^{-4}$ in the dilepton+photon channel and  $[-0.11;0.11] ([-0.084;0.084])$ TeV$^{-4}$, $[-0.092;0.092] ([-0.072;0.072])$ TeV$^{-4}$ in the MET+photon channel with $L_{int}$=1 (3) ab$^{-1}$, respectively. 
\end{abstract}

\maketitle

\section{Introduction}
The gauge boson self-interactions represented by the non-Abelian $SU(2)_L \times U(1)_Y$ gauge group of the electroweak sector in the Standard Model (SM) are of great interest since it provides valuable information to test the predictions at the TeV energy scale. The triple couplings between the photon and Z boson ($Z\gamma\gamma$ and $Z\gamma Z$) vanish at tree level in the SM. Any deviations of these couplings from SM predictions within the experimental precision can give crucial clues about new physics beyond the SM. A method to parameterize these new physics effects at higher energies is Effective Field Theory (EFT) approach which reduces to the SM at low energies. Diboson productions at hadron colliders through EFT framework have been studied in Ref. \cite{Green:2016trm, Dorigo:2018cbl,Baur:2000ae,Mangano:2015ejw,Neubauer:2011zz,Senol:2013lca}. There are also studies (see for example ref.\cite{Frye:2015rba}) about the enhancement for the sensitivity of diboson measurement at the LHC.

The Lagrangian in the framework of an effective field theory for neutral Triple Gauge Couplings (nTGC) imposing local $U(1)_{EM}$ and Lorentz symmetry can be written as \cite{Degrande:2013kka}
\begin{eqnarray}
\mathcal{L}^{nTGC}=\mathcal{L}^{SM}+\sum_i\frac{C_i}{\Lambda^4}(\mathcal{O}_i+\mathcal{O}_i^{\dagger})
\end{eqnarray}
where $i$ runs over the label of the four operators expressed as
\begin{eqnarray}
\mathcal{O}_{BW}&=&iH^{\dagger}B_{\mu\nu}W^{\mu\rho}\{D_{\rho},D^{\nu}\}H\\
\mathcal{O}_{WW}&=&iH^{\dagger}W_{\mu\nu}W^{\mu\rho}\{D_{\rho},D^{\nu}\}H\\
\mathcal{O}_{BB}&=&iH^{\dagger}B_{\mu\nu}B^{\mu\rho}\{D_{\rho},D^{\nu}\}H\\
\mathcal{O}_{\tilde B W}&=&iH^{\dagger} \tilde B_{\mu\nu}W^{\mu\rho}\{D_{\rho},D^{\nu}\}H
\end{eqnarray}
where $\tilde B_{\mu\nu}$ is a dual $B$ strength tensor. The following convention in the definitions of the operators are used: 
\begin{eqnarray}
W_{\mu\nu}&=&\sigma^I(\partial_{\mu} W_{\nu}^I-\partial_{\nu} W_{\mu}^I+g\epsilon_{IJK}W_{\mu}^JW_{\nu}^K)\\
B_{\mu\nu}&=&(\partial_{\mu} B_{\nu}-\partial_{\nu} B_{\mu})
\end{eqnarray}
with $\left<\sigma^I\sigma^J\right>=\delta^{IJ}/2$
and 
\begin{eqnarray}
D_{\mu}\equiv\partial_{\mu}-ig_wW_{\mu}^i\sigma^i-i\frac{g'}{2}B_{\mu}Y
\end{eqnarray}
The coefficients of these four dimension-eight operators describing anomalous Neutral Triple Gauge Couplings (aNTGC) are CP-conserving $C_{\widetilde B W}$ and CP-violating $C_{BB}$, $C_{BW}$, $C_{WW}$. They are related to dimension-six operators aNTGC as described in Ref. \cite{Degrande:2013kka}. The 95\% C.L. current limits on dimension-eight operators converted from coefficients of dimension-six operators for the process $pp\to ZZ\to l^+l^-l'^+l'^-$ at $\sqrt s$=13 TeV and $L_{int}$ =36.1 fb$^{-1}$ from ATLAS collaboration are reported  as \cite{Aaboud:2017rwm}
\begin{eqnarray*}
-5.9~\textrm{TeV}^{-4}<\frac{C_{\widetilde B W}}{\Lambda^4}<5.9~\textrm{TeV}^{-4}\\
-3.0~\textrm{TeV}^{-4}<\frac{C_{ W W}}{\Lambda^4}<3.0~\textrm{TeV}^{-4}\\
-3.3~\textrm{TeV}^{-4}<\frac{C_{B W}}{\Lambda^4}<3.3~\textrm{TeV}^{-4}\\
-2.7~\textrm{TeV}^{-4}<\frac{C_{B B}}{\Lambda^4}<2.8~\textrm{TeV}^{-4}\\
\end{eqnarray*}

Recently, the production of Z boson in association with a high energy photon is studied in the $Z\to\nu\bar \nu$ channel at $\sqrt s$ = 13 TeV with an integrated luminosity of 36.1 fb$^{-1}$ by the ATLAS collaboration and set limits on the EFT parameters at the 95\% C.L. as follows \cite{ATLAS:2018eke}
\begin{eqnarray*}
-1.1~\textrm{TeV}^{-4}<\frac{C_{\widetilde B W}}{\Lambda^4}<1.1~\textrm{TeV}^{-4}\\
-2.3~\textrm{TeV}^{-4}<\frac{C_{ W W}}{\Lambda^4}<2.3~\textrm{TeV}^{-4}\\
-0.65~\textrm{TeV}^{-4}<\frac{C_{B W}}{\Lambda^4}<0.64~\textrm{TeV}^{-4}\\
-0.24~\textrm{TeV}^{-4}<\frac{C_{B B}}{\Lambda^4}<0.24~\textrm{TeV}^{-4}\\
\end{eqnarray*}
In this study, we investigate the constrains on dimension eight operators in the $pp\to l^+l^-\gamma$ and $pp\to \nu\bar \nu \gamma$ processes since photon with high transverse momentum enhance the existence of aNTGCs \cite{Barger:1984yn, Baur:1992cd, Baur:1997kz}. One can expect even further improvements on these bounds with a 100 TeV center of mass energy collider. The Future Circular Collider (FCC) which has the potential to search for a wide parameter range of new physics is the energy frontier collider project currently under consideration \cite{FCC}. One of its options (called FCC-hh) has a design providing proton-proton collisions at the proposed centre-of-mass energy of 100 TeV with peak luminosity of $5\times10^{34}$ $cm^{-2}s^{-1}$ \cite{Mangano:2017tke}.

\begin{figure}
\includegraphics{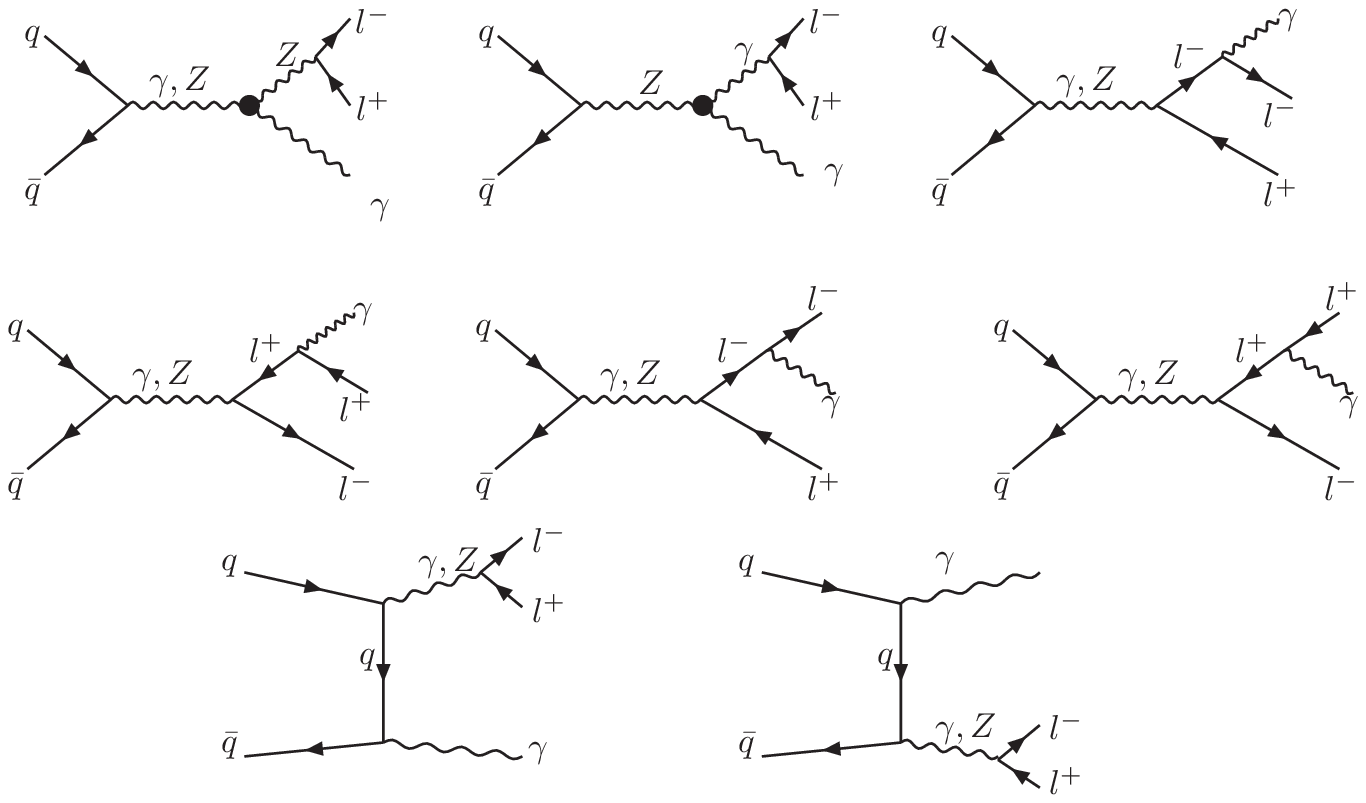}
\caption{ Feynman diagrams for $pp\to l^-l+\gamma$ process contributing in the SM and anomalous $ZZ\gamma$, $Z\gamma\gamma$ vertices.  \label{fd_1}}
\end{figure}

\begin{figure}
\includegraphics{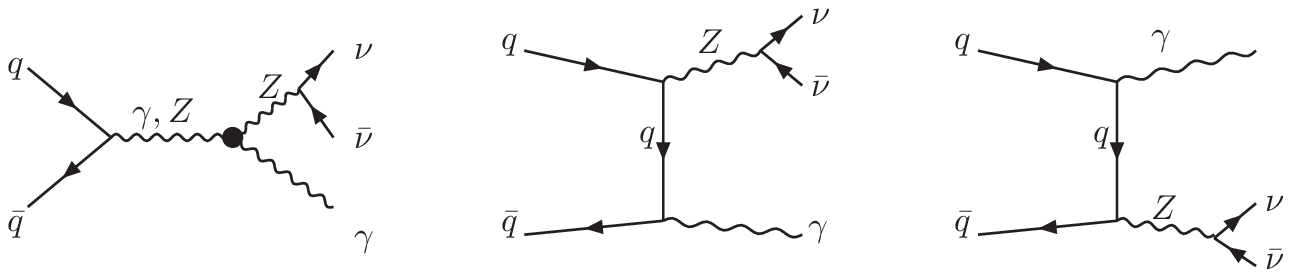}
\caption{ Feynman diagrams for $pp\to \nu\bar \nu\gamma$ process contributing in the SM and anomalous $ZZ\gamma$, $Z\gamma\gamma$ vertices.  \label{fd_2}}
\end{figure}

The tree level Feynman diagrams of the $pp\to l^+l^-\gamma$ process and $pp\to \nu\bar \nu \gamma$ process are shown in Fig.~\ref{fd_1} and Fig.~\ref{fd_2}, respectively. The first two diagrams account for the anomalous $Z\gamma\gamma$ and $ZZ\gamma$ couplings, while the others for SM contributions in Fig.~\ref{fd_1}. In Fig.~\ref{fd_2}, the first diagram consists of anomalous  $Z\gamma\gamma$ and $ZZ\gamma$ couplings and the others comes from SM electroweak processes.  In order to calculate the effects of dimension-eight operators on $pp\to l^+l^-\gamma$ and $pp\to \nu\bar \nu \gamma$ processes, we use \verb|MadGraph5_aMC@NLO| \cite{Alwall:2014hca} after implementation of the operators Eqs.(2)-(5) through Feynrules package \cite{Alloul:2013bka}  as a Universal FeynRules Output (UFO) module \cite{Degrande:2011ua}. In the next section, we give details of the simulation and discuss for the determination of the limits on the dimension-eight $Z\gamma\gamma$ and $ZZ\gamma$ couplings at 95\% C.L.

\section{Analysis and Simulation details}
\begin{figure}
\includegraphics[scale=0.64]{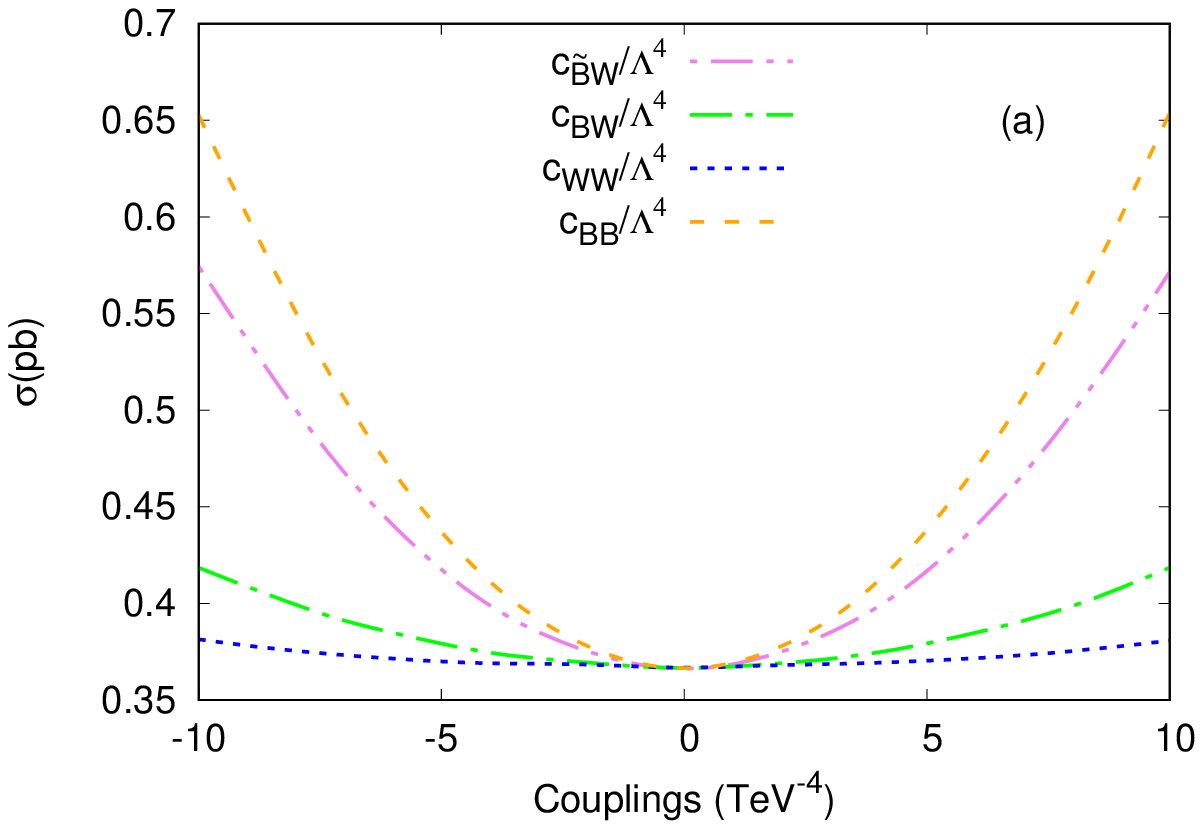} 
\includegraphics[scale=0.64]{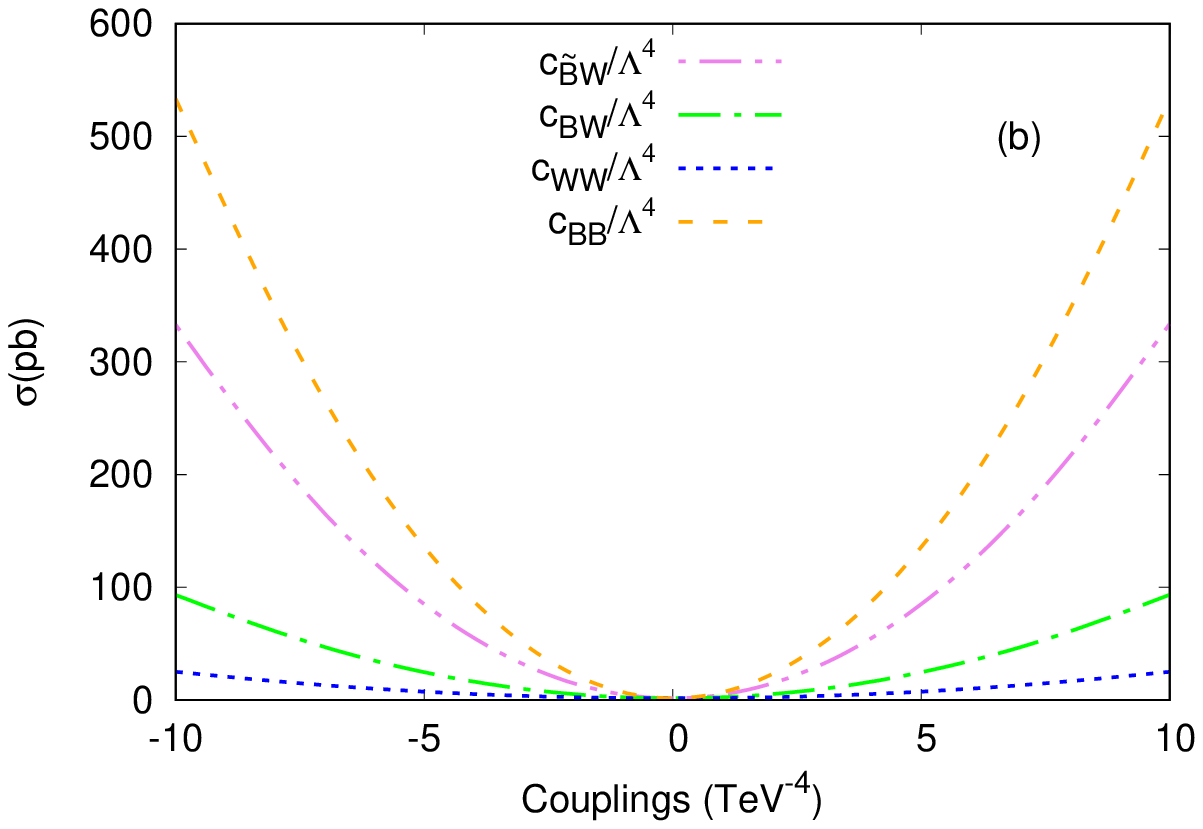} 
\caption{ The signal cross sections of a) $pp\to l^-l^+\gamma$ and b) $pp\to \nu\bar \nu\gamma$ depending on dimension-eight couplings
at FCC-hh.  \label{crosssection} }
\end{figure}

The cross sections of the $pp\to l^+l^-\gamma$  and $pp\to \nu\bar \nu \gamma$ process as a function of mentioned four dimension-eight couplings are shown in Fig.~\ref{crosssection}. In this figure, only one coupling at a time is varied from its SM value. The cross section is calculated with a set of generator level cuts; 

 i) for the process $pp\to l^+l^-\gamma$,  photon transverse momentum $p_T^{\gamma}$>100 GeV and pseudorapidity $|\eta^{\gamma}|<2.5$, charged lepton transverse momentum $p_T^l>20$ GeV and pseudorapidity $|\eta^l|<2.5$.   A charged lepton-photon separation in the pseudorapidity-azimuthal angle plane is defined as follow
\begin{eqnarray}
 \Delta R(l,\gamma)= \left[(\Delta\phi_{l,\gamma}])^2+(\Delta\eta_{l,\gamma}])^2\right]^{1/2}
 \end{eqnarray}
 We also imposed the cuts on $\Delta R (l,\gamma)>0.7$. A large separation cut not only suppress photon radiation from the final state lepton but also impose the lepton-photon separation sharply peaks at small value in radiative Z decays due to the collinear singularity associated with diagrams. The invariant mass of final state charged leptons $m_{ll}$ cut is $m_{ll}>50$ GeV. 
 
ii)  For $pp\to \nu\bar \nu \gamma$ process, the only photon transverse momentum $p_T^{\gamma}$>100 GeV and pseudorapidity $|\eta^l|<2.5$ cuts are applied. 

As it can be seen from Fig.~\ref{crosssection}, deviation from SM value of the anomalous cross section including $C_{\widetilde B W}$ and $C_{BB}$ couplings is larger than that for $C_{BW}$, $C_{WW}$ in the two processes. Therefore, in our analysis we focus on CP-even $C_{\widetilde B W}$  coupling and CP-odd $C_{BB}$ coupling. 
 
 In the analysis, we include effective dimension-eight aNTG couplings and SM contribution as well as interference between effective couplings and SM contributions that lead to $p p \to l^+l^-\gamma$ (Dilepton + photon final state) and $pp\to \nu\bar \nu \gamma$ (MET+photon final state) processes where $l^{\pm}=e^{\pm},\mu^{\pm}$. For the further detailed analysis, the regenerated signal (for $C_{\widetilde B W}/\Lambda^4$=2.0, 4.0, 6.0 TeV$^{-4}$; $C_{B B}/\Lambda^4$=1.0, 2.0, 3.0 TeV$^{-4}$)  and background events at parton level in \verb|MadGraph5_aMC@NLO| applying  pseudo-rapidity $|\eta^{l,\gamma}|<2.5$, and transverse momentum $p_T^{l,\gamma}>20$ GeV  cuts are passed through the Pythia 6 \cite{Sjostrand:2006za} for parton shower and hadronization . The detector responses are taken into account with FCC detector card in \verb|Delphes 3.3.3| \cite{deFavereau:2013fsa} package. Then, all events are analysed by using the ExRootAnalysis utility \cite{exroot} with ROOT \cite{Brun:1997pa}. The kinematical distributions are normalized to the number of expected events which is defined to be the cross section of each processes times integrated luminosity of $L_{int}$=1 ab$^{-1}$. 
 
\begin{figure}
\includegraphics[scale=0.4]{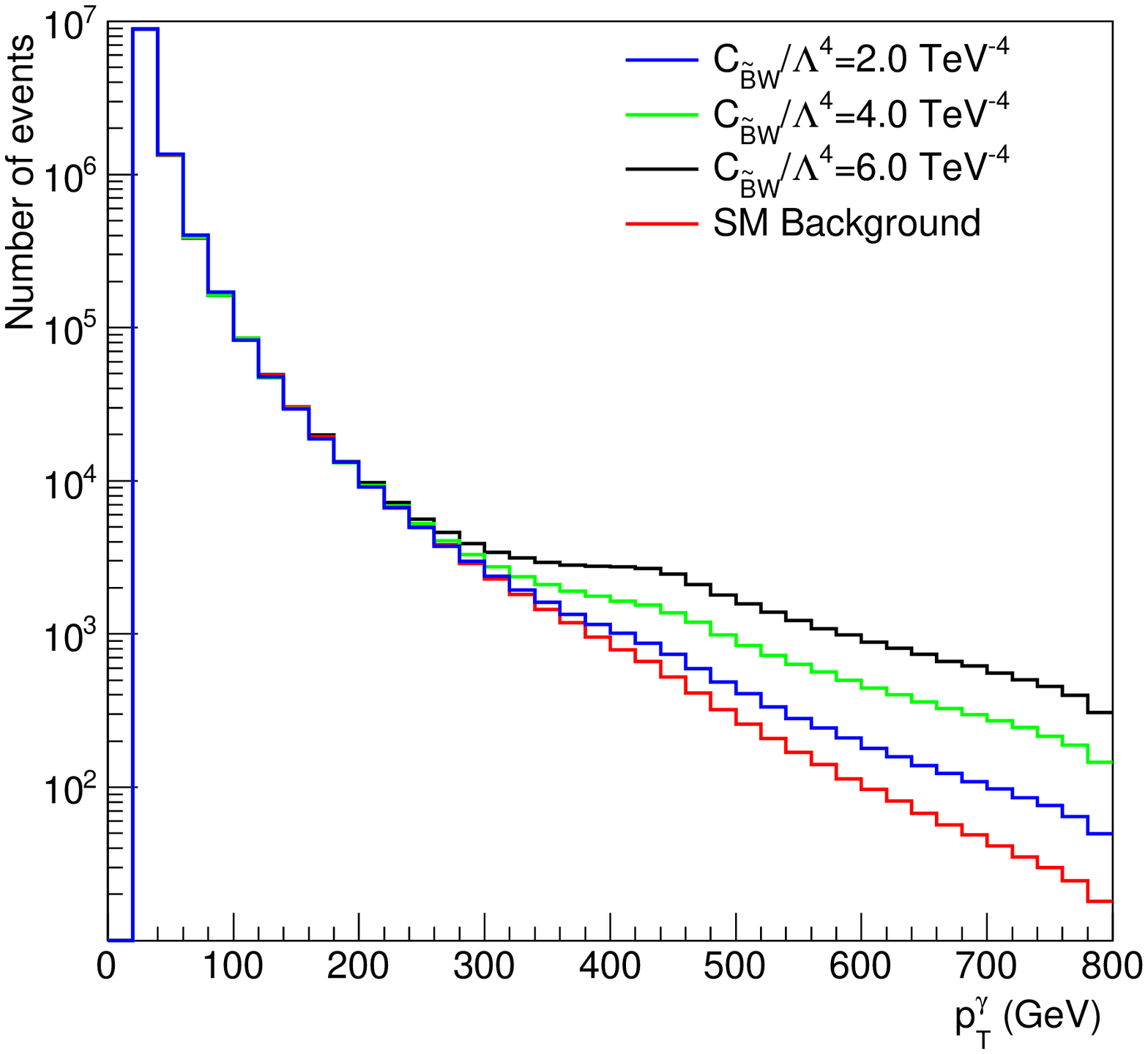} 
\includegraphics[scale=0.4]{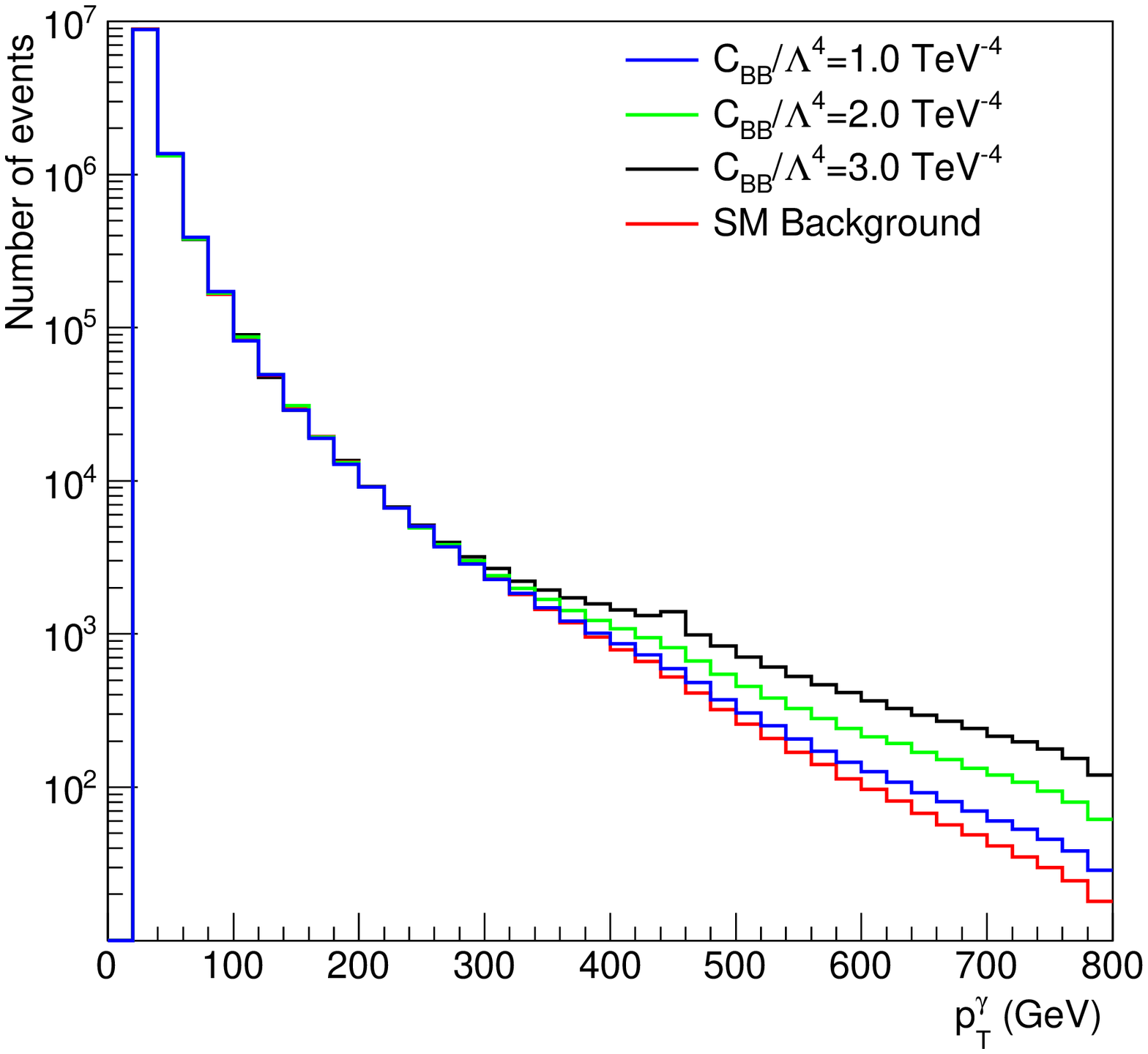} 
\caption{The  $p_T^{\gamma}$ distribution for signal for $C_{\widetilde B W}/\Lambda^4$ (left panel) and $C_{B B}/\Lambda^4$ (right panel) couplings and corresponding SM background of $pp\to l^-l^+\gamma$ process.  \label{ptphoton}}
\end{figure}

\begin{figure}
\includegraphics[scale=0.4]{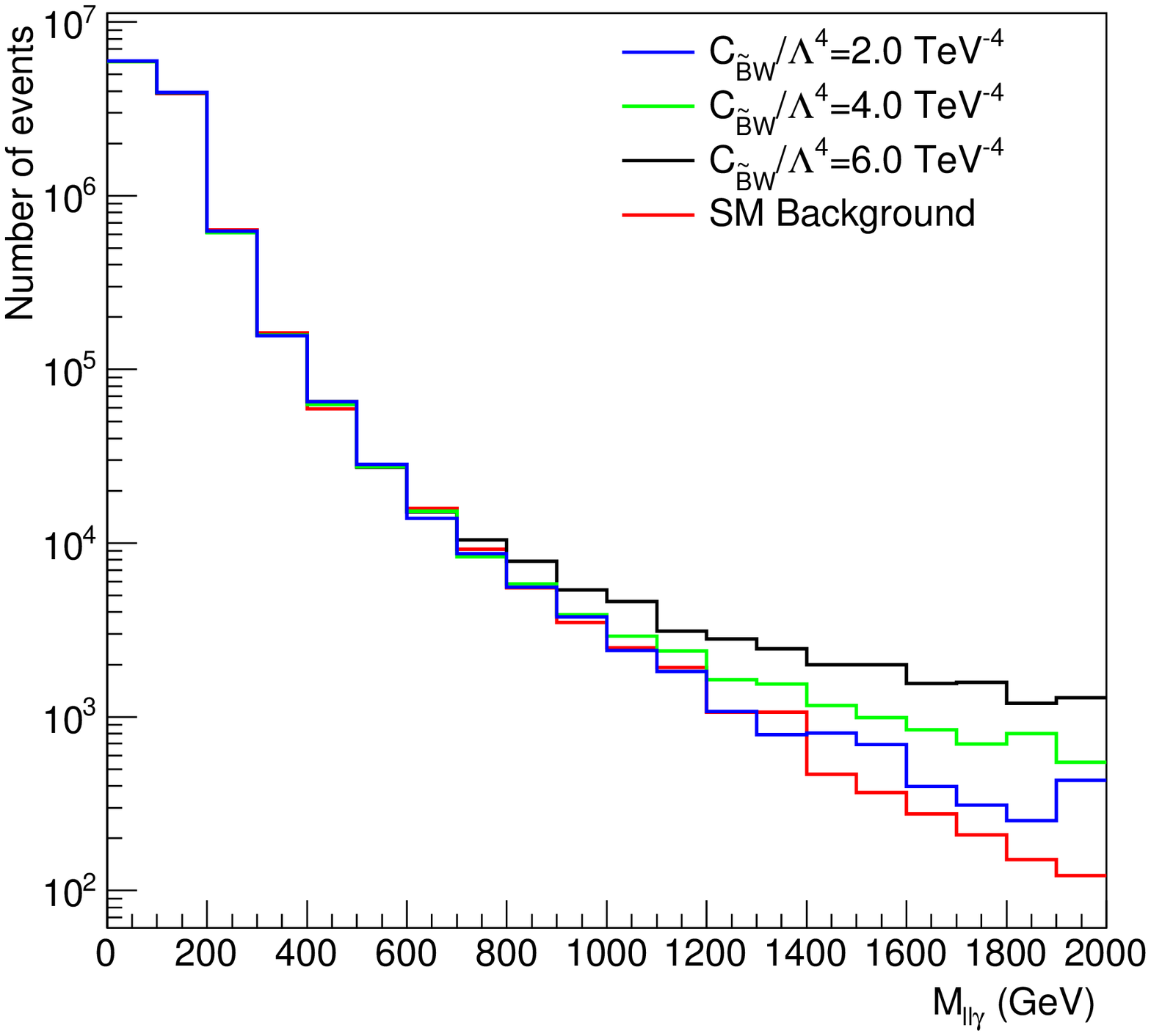} \includegraphics[scale=0.4]{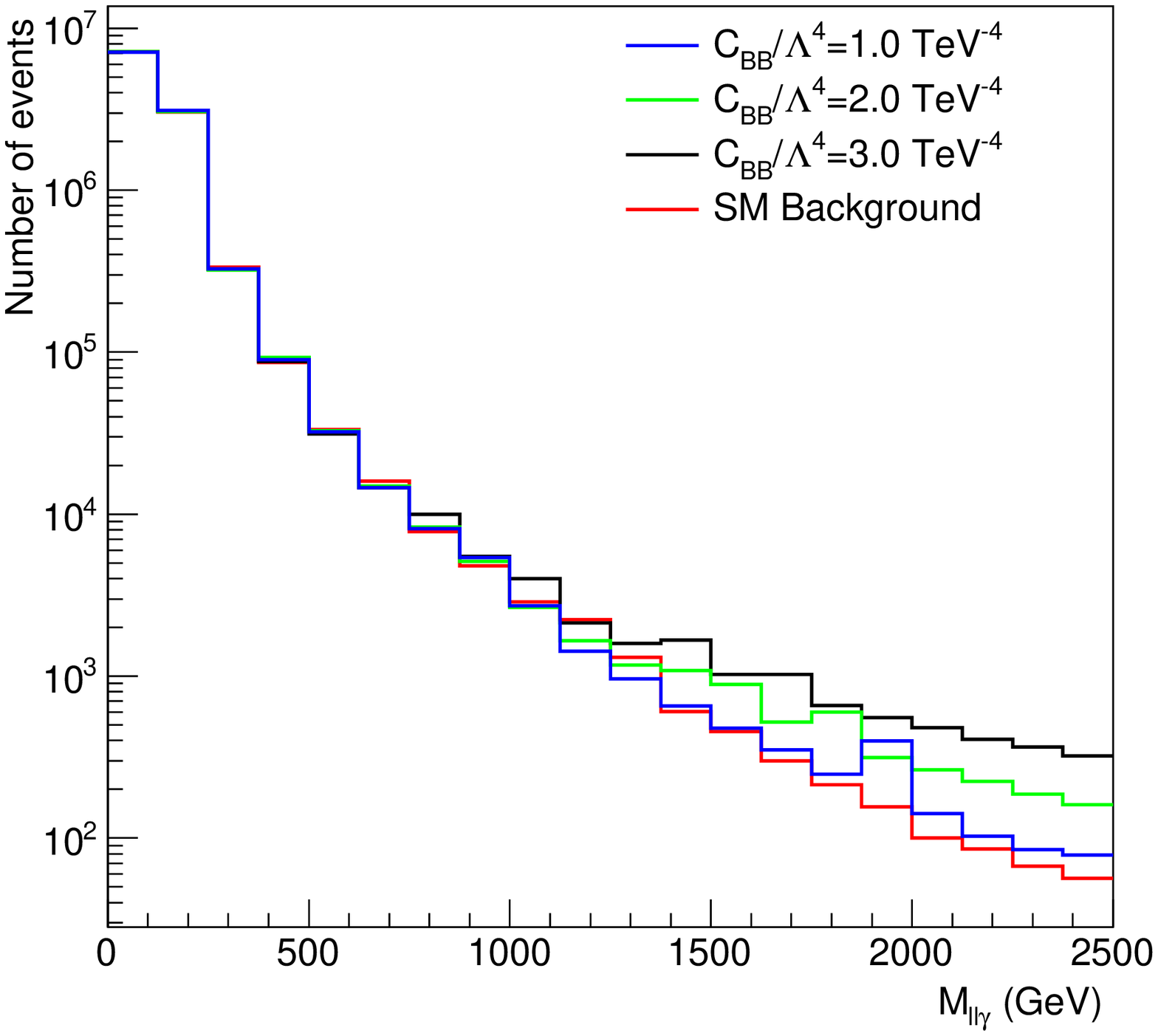} 
\caption{The  invariant mass distribution of $l^+l^-\gamma$ system in $pp\to l^-l^+\gamma$ process for $C_{\widetilde B W}/\Lambda^4$ (left panel) and $C_{B B}/\Lambda^4$ (right panel) couplings and corresponding SM background. \label{minv} }
\end{figure}

\textbf{ $\bullet$ Dilepton+photon analysis}\\
We require  one photon and at least two charged leptons ($l^{\pm}=e^{\pm},\mu^{\pm}$); same flavor but opposite sign. We also require the angular separation $\Delta R(l,\gamma) > 0.7$. In order to define the region for a distinctive signature of the signal, we plot transverse momentum distribution of photon in the final state for $p p \to l^+l^-\gamma$ in Fig.~\ref{ptphoton}. It is clearly seen from Fig.~\ref{ptphoton} that the deviation of the signal from SM background for all couplings starts at about $p_T^{\gamma}$ = 200 GeV. We plot the invariant mass distributions of the $l^+l^-\gamma$ system for signals and SM background in Fig.~\ref{minv}. It shows the deviation from SM background for signal $C_{\widetilde B W}/\Lambda^4$ and $C_{B B}/\Lambda^4$ couplings which appear broader especially at large values of $m_{ll\gamma}$>500 GeV. Therefore, to efficiently separate signal from the SM background, we impose the following cuts; $p_T^{\gamma}$ > 400 (300) GeV, $m_{ll\gamma}$>500 GeV and $m_{ll}>50$ GeV in addition to above mentioned cuts. 
 
\begin{figure}
\includegraphics[scale=0.4]{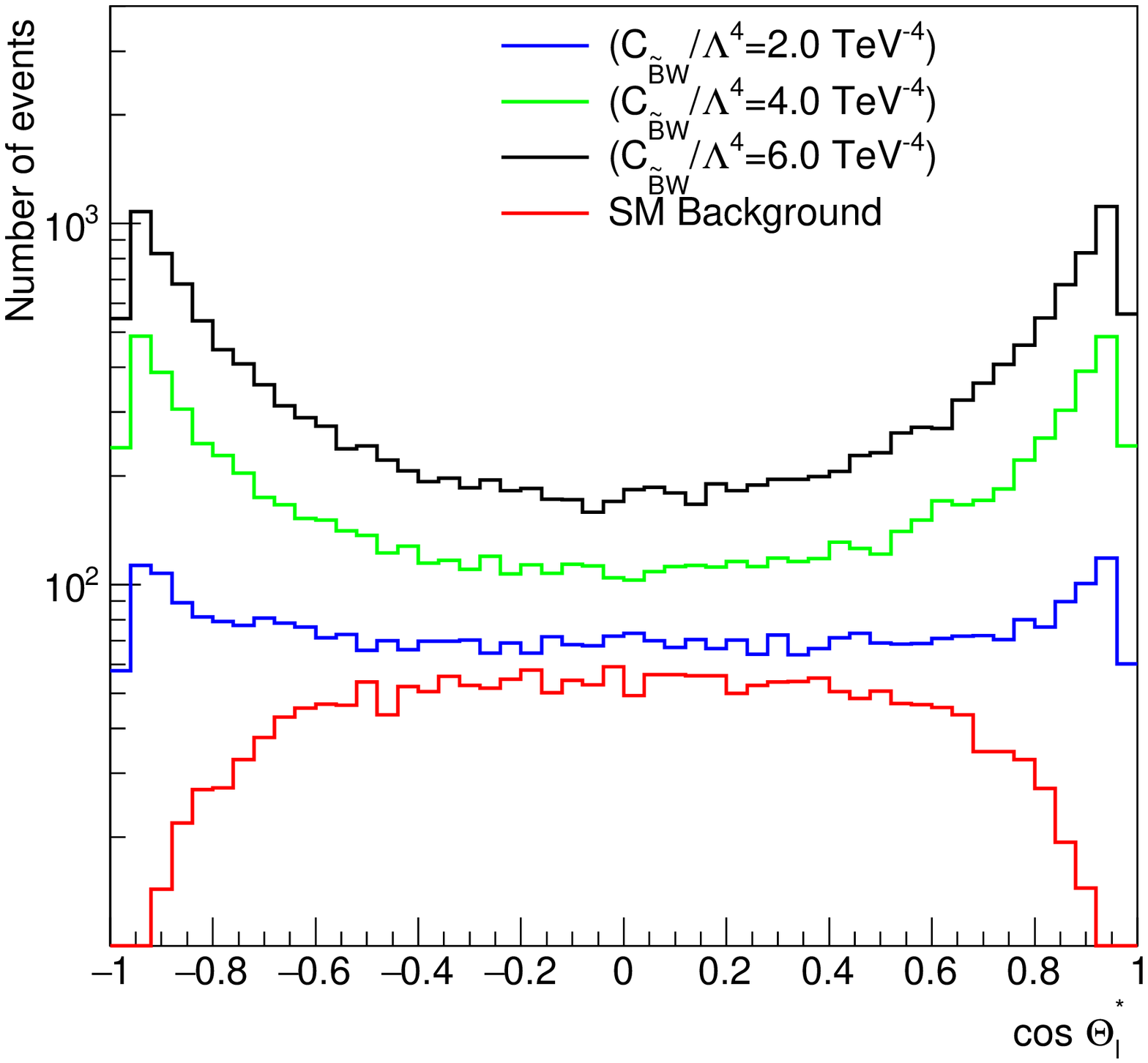} 
\includegraphics[scale=0.4]{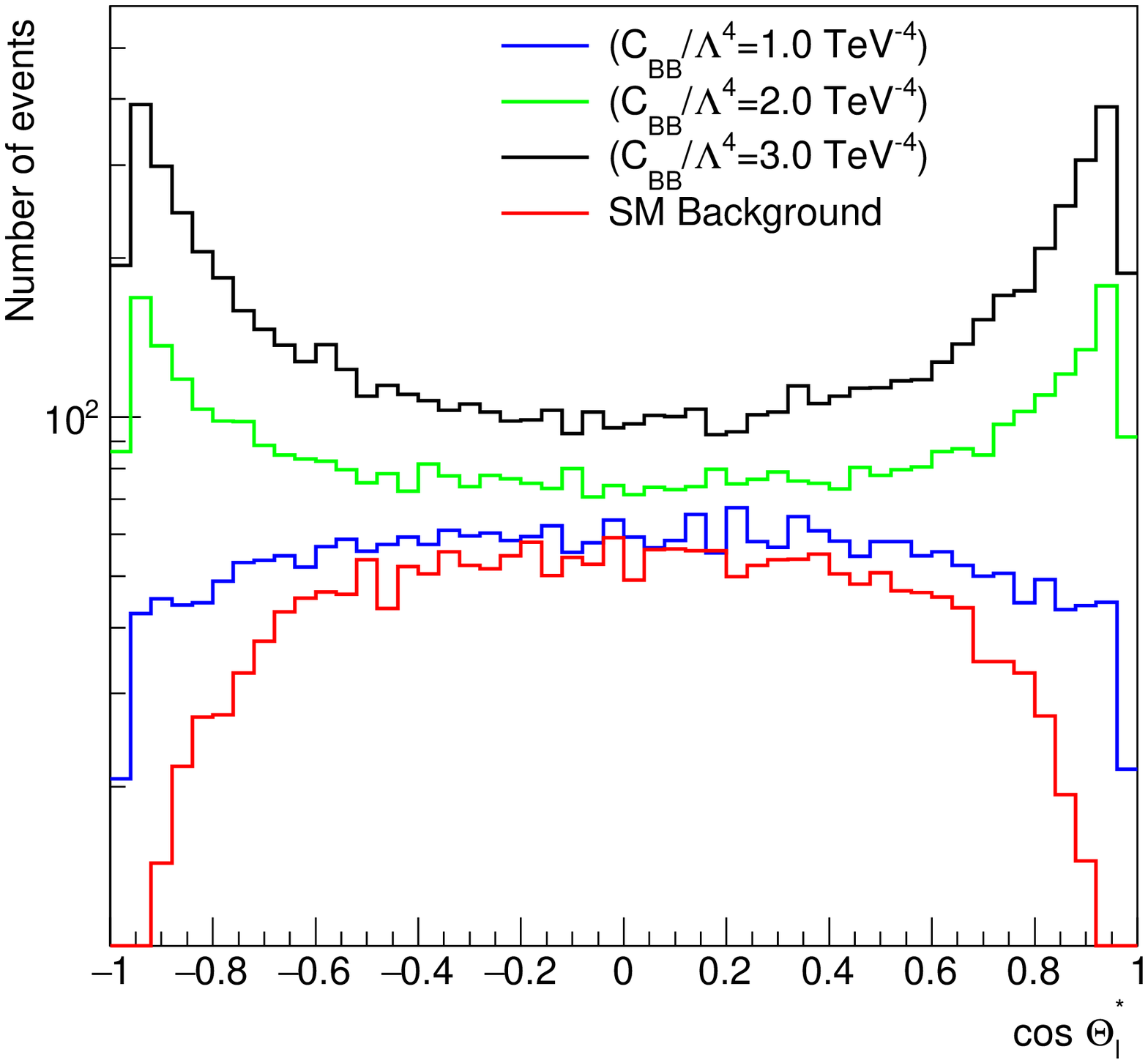} 
\caption{$\cos \Theta^*_l$  distributions for $C_{\widetilde B W}/\Lambda^4$ (left panel) and $C_{B B}/\Lambda^4$ (right panel) and SM background of the $pp\to l^-l^+\gamma$ process. \label{costheta}}
\end{figure}

The angular distribution of final state particles of signal and background processes are used effectively to find attainable limits on effective dimension-eight aNTG couplings since the shape of signal is different from the background. Fig.~\ref{costheta} shows $\cos \Theta^*_l$ distributions of signal for $C_{\widetilde B W}/\Lambda^4$ (left panel), $C_{B B}/\Lambda^4$ (right panel) couplings and SM background. Here, $\cos \Theta^*_l$ is the polar angle in the $l^+l^-$ rest frame with respect to the $l^+l^-$ direction in the $l^+l^-\gamma$ rest frame.  

\textbf{ $\bullet$ MET+photon analysis}\\
Because the neutrinos escape from detection, the invariant mass of final state in this channel cannot be reconstructed. We can use photon $p_T$ distribution and missing transverse energy distribution for the analysis of MET+photon channel. These discriminators are used to search for sensitivity to nonstandard $Z \gamma V$ ($V=\gamma, Z$) couplings. This analysis has the potential advantage since $Z\to\nu\nu$ branching ratio is larger than $Z\to l^+l^-$ ratio. The final state bremsstrahlung and virtual photon do not contribute for the MET+photon channel. The cross section (for MET+photon channel) is about a factor of 3 larger than that of dilepton+photon channel for the standard couplings. In this analysis, we include the interference between signal and background processes. 

\begin{figure}
\includegraphics[scale=0.4]{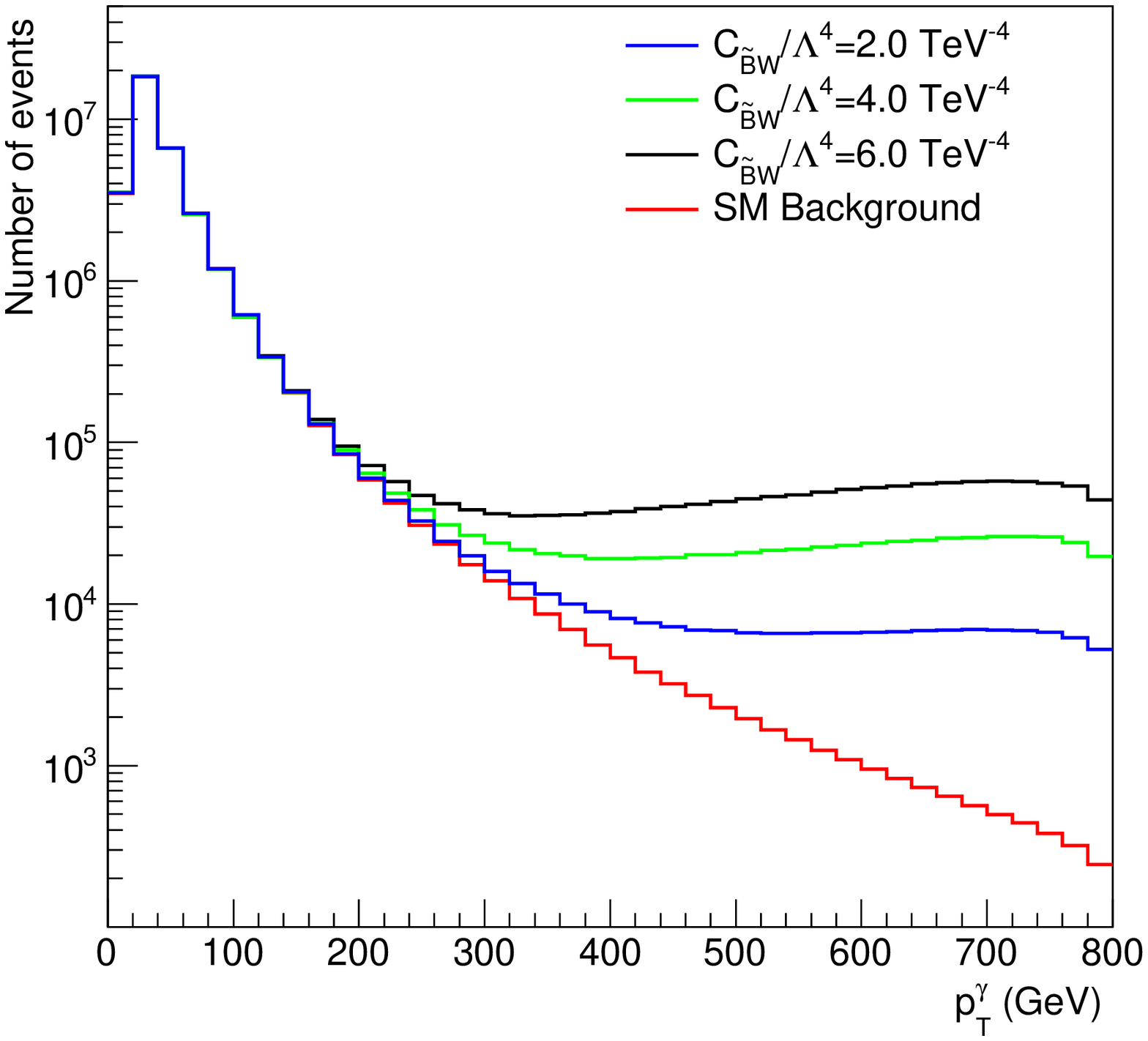} 
\includegraphics[scale=0.4]{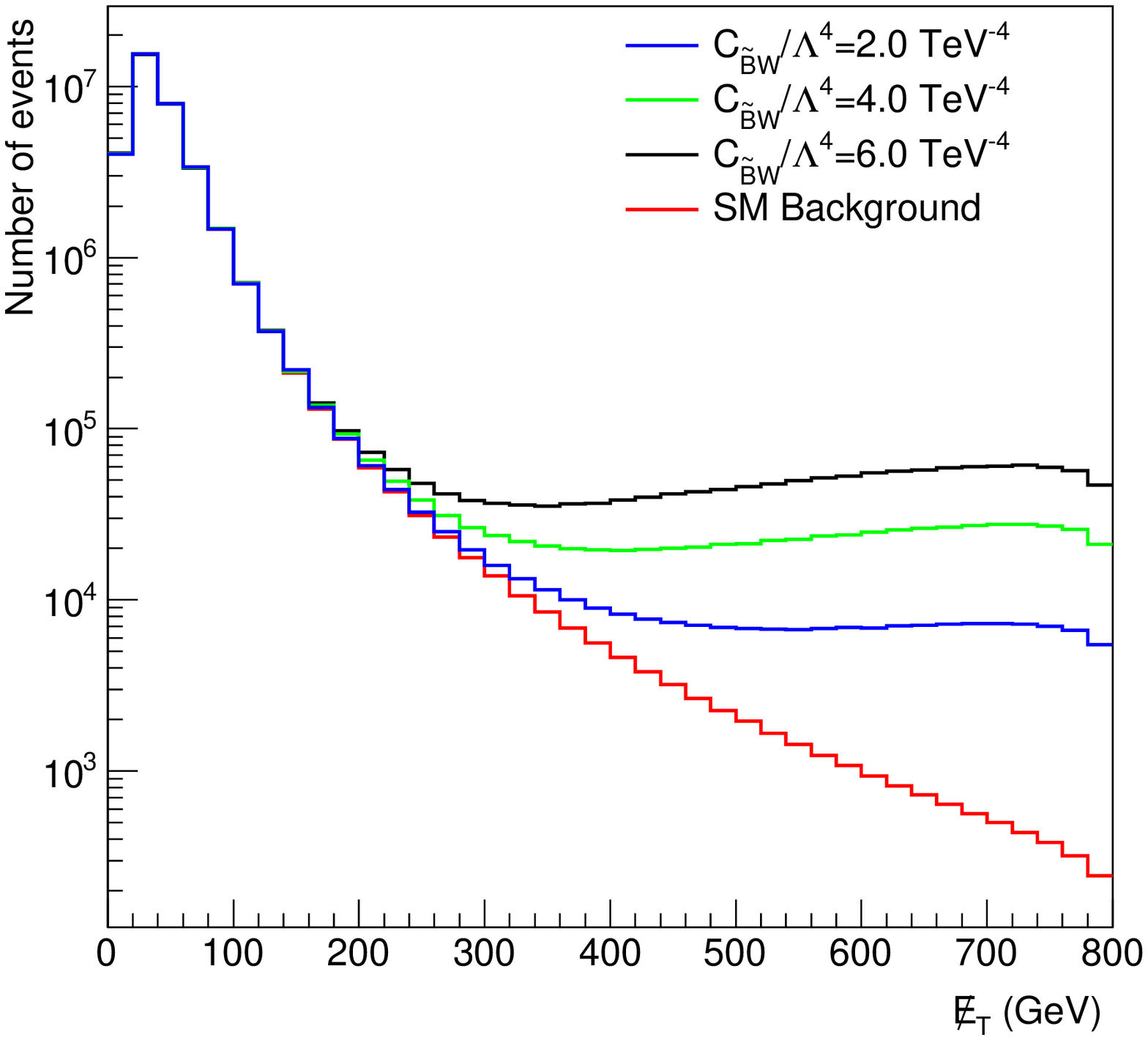} 
\caption{The $p_T^{\gamma}$ distribution (left panel) and MET distribution (right panel) for signal ($C_{\widetilde B W}/\Lambda^4$)  and corresponding SM background of $pp\to \nu\bar \nu \gamma$ process.  \label{pt_met}}
\end{figure}

We select the events with a photon and a large missing transverse energy (MET). The $p_T$ distribution of the photon and MET distribution are given in Fig.~\ref{pt_met}. As can be seen from Fig.~\ref{pt_met}, the discrimination of signal from the SM background for all couplings well appears for $p_T^{\gamma}$ > 300 GeV and MET$>300$ GeV. Therefore, we impose the following cuts for two parameters in the analysis; $p_T^{\gamma}$ > 400 (300) GeV,  and MET > 400 GeV. 

\section{Results of the Analysis}
In the analysis, we use the angular distribution ($\cos \Theta^*_l$)  in the dilepton+photon channel and transverse momentum distribution of the photon ($p_T^{\gamma}$) in MET+photon channel for the signal and background.
To obtain 95\% C.L. limits on the couplings, we apply $\chi^{2}$ criterion with and without a systematic error. The $\chi^{2}$ function is defined as follows
\begin{eqnarray}
\chi^{2} =\sum_i^{n_{bins}}\left(\frac{N_{i}^{NP}-N_{i}^{B}}{N_{i}^{B}\Delta_i}\right)^{2}
\end{eqnarray}
where $N_i^{NP}$ is the total number of events in the existence of effective couplings, $N_i^B$ is total number of events of the corresponding SM backgrounds in $i$th bin of the $\cos \Theta^*_l$ and $p_T^{\gamma}$ distributions, $\Delta_i=\sqrt{\delta_{sys}^2+\frac{1}{N_i^B}}$ is the combined systematic ($\delta_{sys}$) and statistical errors in each bin. For the analysis of dilepton+photon channel, the number of signal events and one-parameter $\chi^{2}$ results for $C_{\widetilde B W}/\Lambda^4$=2.0, 4.0, 6.0 TeV$^{-4}$  and $C_{B B}/\Lambda^4$=1.0, 2.0, 3.0 TeV$^{-4}$ are given in Table~\ref{tab1} and Table~\ref{tab2}, respectively. In these tables, only one coupling at a time is varied from its SM value. We also present numerical results taking into account systematic errors, $\delta_{sys}=5\%$ and $\delta_{sys}=10\%$  for $p_T^{\gamma}$ > 400 GeV ( $p_T^{\gamma}$ > 300 GeV in the parenthesis) at an integrated luminosity of 1 ab$^{-1}$. Here, the number of SM background events is 1098.57 (2076.41). For the analysis of MET+photon channel, the number of signal events and one-parameter $\chi^{2}$ results for the same coupling values of $C_{\widetilde B W}/\Lambda^4$ and $C_{B B}/\Lambda^4$ are given in Table~\ref{tab3} and Table~\ref{tab4}, respectively. For the same $p_T^{\gamma}$ cuts, the number of SM background events is 22761 (27213).

\begin{figure}
\includegraphics[scale=0.4]{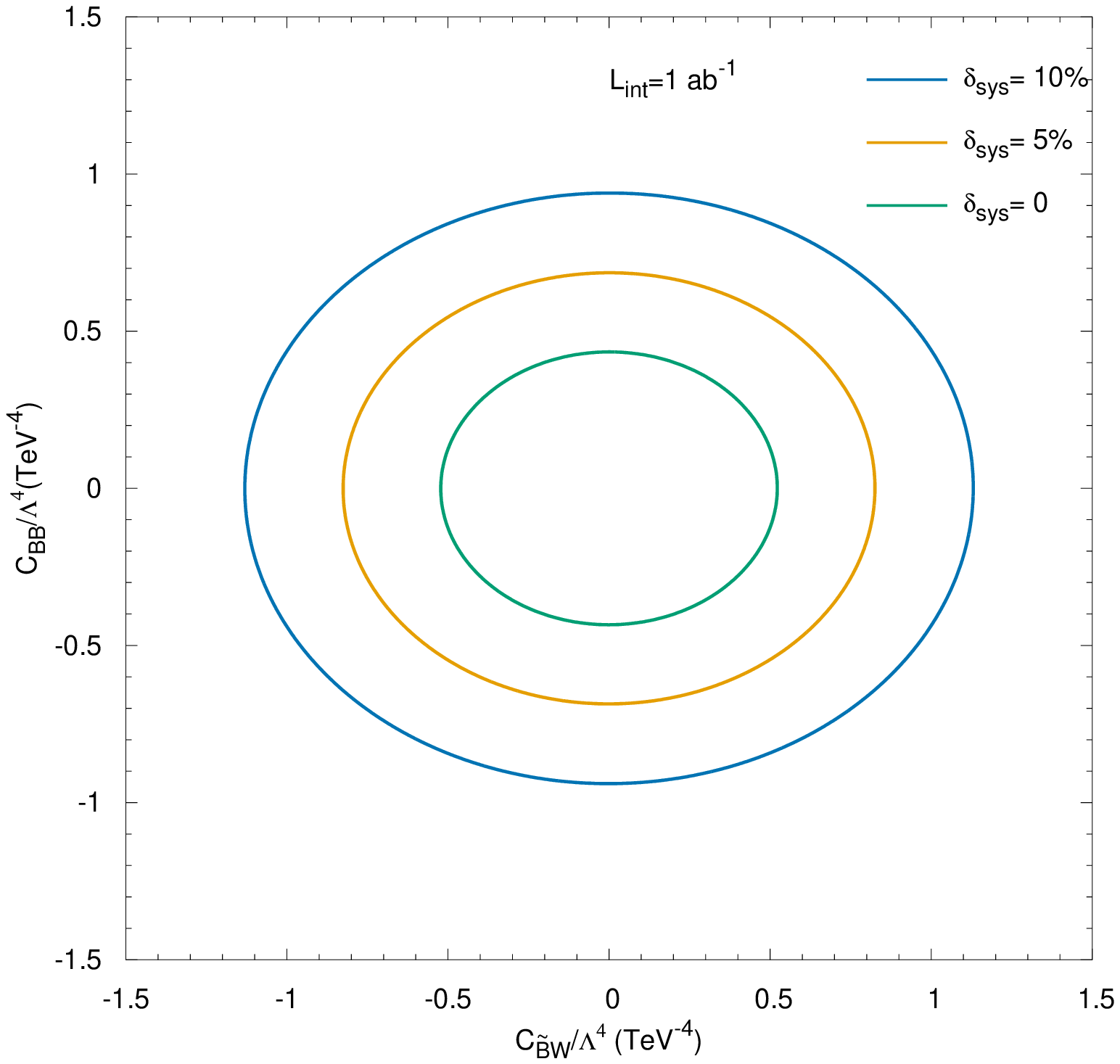} 
\includegraphics[scale=0.4]{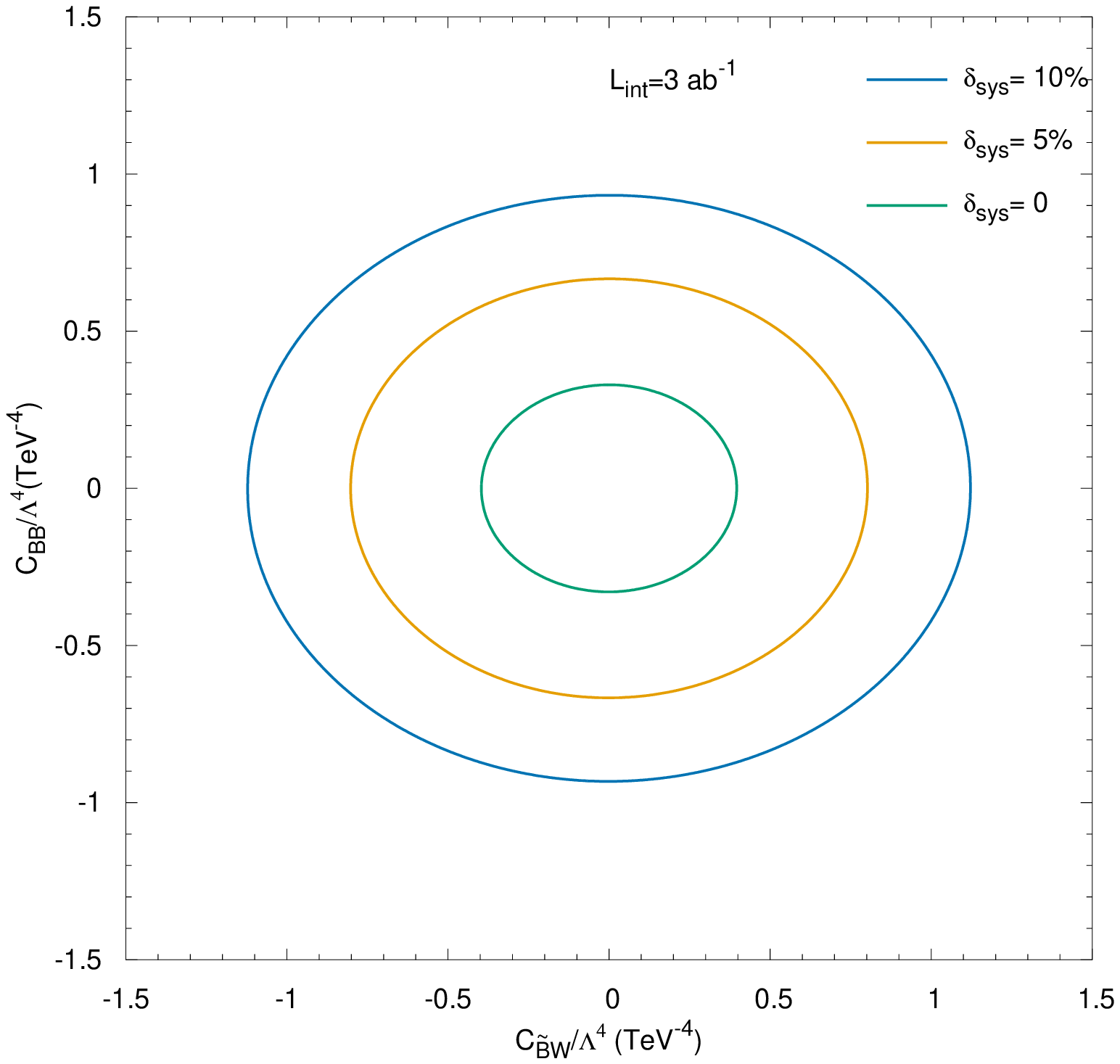}  
\caption{ Two-dimensional 95\% C.L. intervals in plane for $C_{\widetilde B W}/\Lambda^4$ and $C_{B B}/\Lambda^4$ with  taking $\delta_{sys}$=0, $\delta_{sys}$=5\% and $\delta_{sys}$=\%10 of systematic errors at $L_{int}$=1 and 3 ab$^{-1}$ for $pp\to l^-l^+\gamma$ process.\label{twod}}
\end{figure}

\begin{figure}
\includegraphics[scale=0.4]{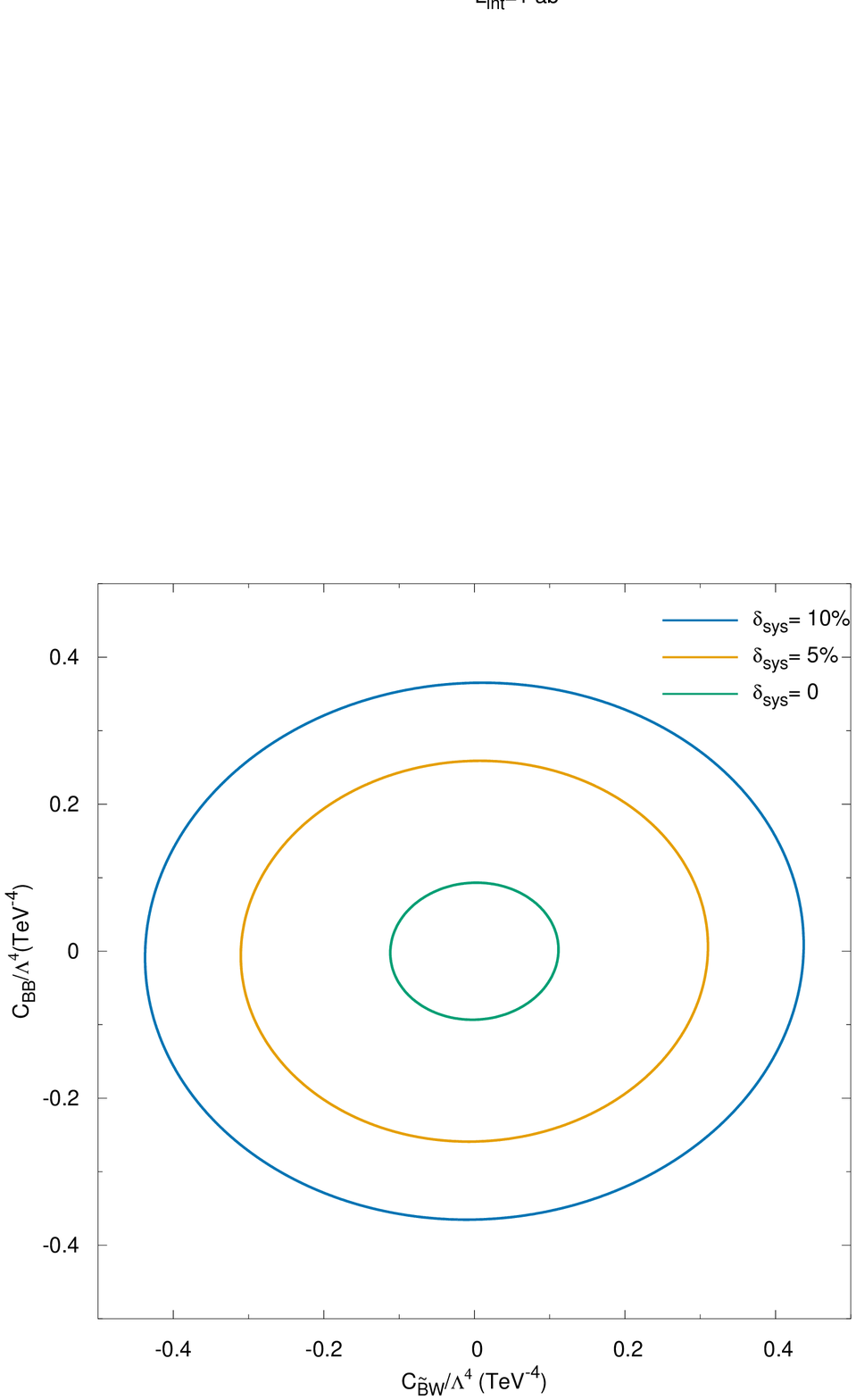} 
\includegraphics[scale=0.4]{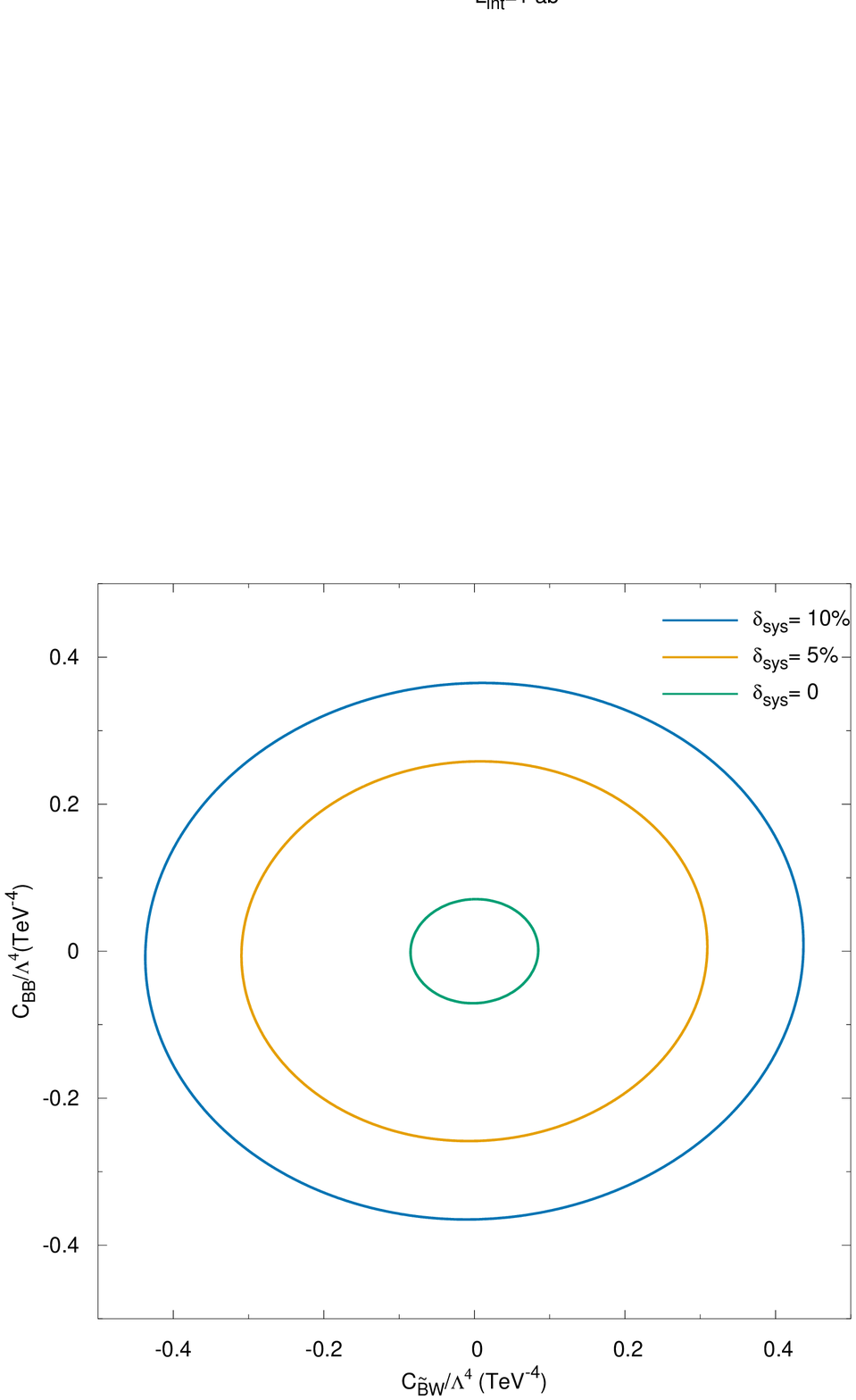}  
\caption{ Two-dimensional 95\% C.L. intervals in plane for $C_{\widetilde B W}/\Lambda^4$ and $C_{B B}/\Lambda^4$ with  taking $\delta_{sys}$=0, $\delta_{sys}$=5\% and $\delta_{sys}$=\%10 of systematic errors at $L_{int}$=1 and 3 ab$^{-1}$ for the $pp\to \nu\bar \nu\gamma$ process.\label{twod_nna}}
\end{figure}

The 95\% C.L. intervals are obtained by allowing pairs of $C_{\widetilde B W}/\Lambda^4$ and $C_{B B}/\Lambda^4$ couplings to vary, while setting the others to zero. The results from two-parameter $\chi^{2}$ analysis of the $C_{\widetilde B W}/\Lambda^4$ and $C_{B B}/\Lambda^4$ couplings at $L_{int}$=1 and 3 ab$^{-1}$ without and with systematic errors (5\% and 10\%)  are shown in Fig.~\ref{twod} (Fig.~\ref{twod_nna}) for dilepton+photon channel (for MET+photon channel). The results on the bounds of dimension-eight aNTG couplings at 95\% C.L. for $L_{int}=1$ and 3 ab$^{-1}$ from these figures are given in Table~\ref{tab5} and Table\ref{tab6} for dilepton+photon channel and in Table~\ref{tab7} and Table~\ref{tab8} for MET+photon channel.  Our limits without systematic error are about one order of magnitude better than the ATLAS bounds on $C_{\widetilde B W}/\Lambda^4$ and $C_{B B}/\Lambda^4$ couplings \cite{Aaboud:2017rwm} and \cite{ATLAS:2018eke}. Even including 5\% systematic error, we obtain limits for $L_{int}$=1 ab$^{-1}$, about three times better than current experimental limits. Without systematic errors, the integrated luminosity has the effect on the bounds of couplings, however the injection of a systematic error $\delta_{sys}=5\%$ prevent sensible changes of the coupling bounds when the luminosity increase. As a result, we find that our bounds on the couplings are systematically limited.

In the effective field theory (EFT) approach, the effective Lagrangian is written to extend the SM Lagrangian by a set of higher dimensional operators, with the assumption that produces low energy limit of a more fundamental description. In the expansion Eq. (1), the coefficients has a scaling to hold UV completion and admit perturbative expansion
in its couplings $C_{i}^{(8)}\sim g^{2}/\Lambda^{4}$ for the operators made of four fields, and this issue has been explained in Ref. \cite{Degrande:2013kka,Contino:2016jqw}. An additional suppression factor may appear with respect to the scaling if an operator is generated at loop level. If no perturbative expansion is possible in the UV theory because it is maximally strongly coupled, then this scaling gives a correct estimate of the size of the effective coefficients by setting coupling $\sim4\pi$. In order to check the validity regime of the EFT, we need the minimum coupling value of the coefficients to put the operator scale $\Lambda$ beyond the reach of the kinematical range of the distributions in order for
the EFT approach not to break down. The coefficients of the dimension-eight operators could be related to the new physics characteristic scale $\Lambda$ via $\bar{C}\sim g^{*2}v^{2}\hat{s}_{max}/\Lambda^{4}$ where $g^{*}$ is the coupling constant of the heavy degrees of freedom with the SM particles. An upper bound can be put on the new physics
scale $\Lambda$ using the fact that the underlying theory is strongly coupled by setting $g^{*}=4\pi$. Assuming $\bar{C}=O(1)$, we find $\Lambda<\sqrt{4\pi v\sqrt{\hat{s}_{max}}}\sim17.5$ TeV. This upper bound is not violated in this analysis as we have $M_{ll\gamma}<2.5$ TeV for the kinematic range of invariant mass distributions.

\section{Conclusions}
We have investigated the effects of dimension-eight operators giving rise to anomalous neutral triple gauge boson interactions of $Z\gamma\gamma$ and $Z\gamma Z$ vertices in $l^-l^+\gamma$   and  $\nu\bar \nu \gamma$ production at FCC-hh. Since $pp\to l^-l^+\gamma$ and $pp\to \nu\bar \nu \gamma$ processes are sensitive to transverse momentum of the final state photon, we use this as a tool to probe the sensitivity of $C_{\widetilde B W}/\Lambda^4$ and $C_{B B}/\Lambda^4$ couplings. Invariant mass of $l^-l^+\gamma$ and angular distribution of charged lepton in the rest frame of $l^-l^+$ with realistic detector effects are also considered in the analysis. Assuming that only one dimension-eight operator is non-zero at a time we obtain at 95\% C.L. sensitivity limits in the dilepton+photon channel for $C_{\widetilde B W}/\Lambda^4$ and $C_{B B}/\Lambda^4$ couplings without systematic errors are $[-0.52;0.52] ([-0.40;0.40] )$ TeV$^{-4}$, $[-0.43;0.43] ([-0.33;0.33])$ TeV$^{-4}$ for $L_{int}$=1 (3) ab$^{-1}$, respectively. For MET+photon channel, we obtain  the bounds on the couplings $[-0.11;0.11] ([-0.084;0.084] )$ TeV$^{-4}$, $[-0.092;0.092] ([-0.072;0.072])$ TeV$^{-4}$ for $L_{int}$=1 (3) ab$^{-1}$, respectively.  We conclude that the future 100 TeV circular hadron collider will be able to provide limits on the $Z\gamma\gamma$ and $Z\gamma Z$  dimension-eight couplings which are about one order of magnitude better than those expected from ATLAS collaboration for the process $pp\to ZZ\to l^+l^-l'^+l'^-$ and $pp\to Z\gamma\to\nu\bar \nu \gamma$ at $\sqrt s$=13 TeV and $L_{int}$ =36.1 fb$^{-1}$.  Even with 5\% systematic errors, the obtained bounds for FCC-hh are about three times better than current LHC results. The result of this study on the bounds of aNTG couplings would benefit from the high luminosity when the systematic uncertainties are well reduced below 5\%. The future circular hadron collider with the high center of mass energy (100 TeV) and integrated luminosity (3 ab$^{-1}$) provides stronger limits than the current experimental limits for aNTG couplings.
\begin{acknowledgments}
This work partially supported by the Bolu Abant Izzet Baysal University Scientific Research Projects under the Project no: 2018.03.02.1286.
\end{acknowledgments}

\newpage
 \begin{table}
\caption{The obtained number of signal events and $\chi^2$ results for various coupling value of $C_{\widetilde B W}/\Lambda^4$ after applied kinematic cuts presented in the text using $\cos \Theta^*_l$  distributions of the $pp\to l^-l^+\gamma$ process with $L_{int}=1$ ab$^{-1}$ ( $p_T^{\gamma}$ > 300 GeV in the parenthesis). \label{tab1}}
\begin{ruledtabular}
\begin{tabular}{ccccc}
$C_{\widetilde B W}/\Lambda^4$ (TeV$^{-4}$) &Number of events&$\chi^2(\delta_{sys}=0)$&$\chi^2(\delta_{sys}=5\%)$&$\chi^2(\delta_{sys}=10\%)$ \\ \hline 
2.0 &  3716.56 (7776.9) & 1295.55 (759.12) & 209.26(49.80)&59.53 (13.09) \\
4.0 & 8770.83  (14078.8)& 21583.05 (12330.53) &3486.18(808.92) & 991.68 (212.70)\\
6.0 &  17090.2 (24396.2 ) &108559.43 (61371.67) & 17534.97(4026.18)&4988.00 (1058.63)\\
\end{tabular}
\end{ruledtabular}
\end{table}
 \begin{table}
\caption{The obtained number of signal events and $\chi^2$ results for various coupling value of $C_{BB}/\Lambda^4$ after applied kinematic cuts presented in the text using $\cos \Theta^*_l$  distributions of the $pp\to l^-l^+\gamma$ process with $L_{int}=1$ ab$^{-1}$ ( $p_T^{\gamma}$ > 300 GeV in the parenthesis).  \label{tab2}}\begin{ruledtabular}
\begin{tabular}{ccccc}
$C_{B B}/\Lambda^4$ (TeV$^{-4}$) &Number of events&$\chi^2(\delta_{sys}=0)$&$\chi^2(\delta_{sys}=5\%)$&$\chi^2(\delta_{sys}=10\%)$ \\ \hline 
1.0 & 2686.51 (6469.10) & 179.26 (104.57) & 28.96 (6.86)&8.24 (1.80)\\
2.0 &   4448.03 (8648.37)&  2708.80 (1528.64)& 437.54 (100.28) &124.46 (26.37)\\
3.0 &  7376.82 (12276.60) &13530.25 (7597.99)  &2185.46 (498.45)&621.68 (131.06)\\
\end{tabular}
\end{ruledtabular}
\end{table}

 \begin{table}
\caption{The obtained number of signal events and $\chi^2$ results for various coupling value of $C_{\widetilde B W}/\Lambda^4$ after applied kinematic cuts presented in the text using $p_T^{\gamma}$ distributions of the $pp\to \nu\bar \nu \gamma$ process  with $L_{int}=1$ ab$^{-1}$ ($p_T^{\gamma}$ > 300 GeV in the parenthesis). \label{tab3}}
\begin{ruledtabular}
\begin{tabular}{ccccc}
$C_{\widetilde B W}/\Lambda^4$ (TeV$^{-4}$) &Number of events&$\chi^2(\delta_{sys}=0)$&$\chi^2(\delta_{sys}=5\%)$&$\chi^2(\delta_{sys}=10\%)$ \\ \hline 
2.0 &  156408 (213103) &  2746636(1755835) & 7444(1553)&1880(390) \\
4.0 &  522497 (621017)& 8310925 (4215077) &112040(23413) & 28296 (5878)\\
6.0 &  1119090 (1292130) &40575745 (20802139) & 547003(115549)&138148 (29008)\\
\end{tabular}
\end{ruledtabular}
\end{table}

 \begin{table}
\caption{The obtained number of signal events and $\chi^2$ results for various coupling value of $C_{BB}/\Lambda^4$ after applied kinematic cuts presented in the text using $p_T^{\gamma}$ distributions of the $pp\to \nu\bar \nu \gamma$ process  with $L_{int}=1$ ab$^{-1}$ ($p_T^{\gamma}$ > 300 GeV in the parenthesis).   \label{tab4}}
\begin{ruledtabular}
\begin{tabular}{ccccc}
$C_{BB}/\Lambda^4$ (TeV$^{-4}$) &Number of events&$\chi^2(\delta_{sys}=0)$&$\chi^2(\delta_{sys}=5\%)$&$\chi^2(\delta_{sys}=10\%)$ \\ \hline 
1.0 &  67598 (74090) &  88329(80749) & 1525(1170)&386(296) \\
2.0 &  194248 (205373)& 1292061 (1166391) &22315(16896) & 5652 (4270)\\
3.0 &  418947 (437906) & 6896322 (6198094) & 119105(89785)&30167 (22693)\\
\end{tabular}
\end{ruledtabular}
\end{table}
 \begin{table}
\caption{ Obtained limits  on $C_{\widetilde BW}/\Lambda^4$ and $C_{BB}/\Lambda^4$ at 95\% C.L. with $L_{int}=1$ ab$^{-1}$ by assuming a non-zero dimension-eight operator at a time for $pp\to l^-l^+\gamma$ process.  \label{tab5}}\begin{ruledtabular}
\begin{tabular}{lccc}
Couplings (TeV$^{-4}$) &$\delta_{sys}=0$&$\delta_{sys}=5\%$&$\delta_{sys}=10\%$ \\ \hline 
$C_{\widetilde B W}/\Lambda^4$  & [-0.52;0.52] & [-0.83;0.83] & [-1.13;1.13]\\
$C_{B B}/\Lambda^4$  &   [-0.43;0.43]&  [-0.68;0.68] & [-0.94;0.94] \\
\end{tabular}
\end{ruledtabular}
\end{table}
 \begin{table}
\caption{  Obtained limits on $C_{\widetilde B W}/\Lambda^4$ and $C_{ B B}/\Lambda^4$ at 95\% C.L. with $L_{int}=3$ ab$^{-1}$ by assuming a non-zero dimension-eight operator at a time for  $pp\to l^-l^+\gamma$ process .  \label{tab6}}\begin{ruledtabular}
\begin{tabular}{lccc}
Couplings (TeV$^{-4}$) &$\delta_{sys}=0$&$\delta_{sys}=5\%$&$\delta_{sys}=10\%$ \\ \hline 
$C_{\widetilde B W}/\Lambda^4$  &[-0.52;0.52] & [-0.83;0.83] & [-1.13;1.13]\\
$C_{B B}/\Lambda^4$  &   [-0.33;0.33]&  [-0.67;0.67] & [-0.93;0.93] \\
\end{tabular} 
\end{ruledtabular}
\end{table}

 \begin{table}
\caption{ Obtained limits on $C_{\widetilde B W}/\Lambda^4$ and $C_{ B B}/\Lambda^4$ at 95\% C.L. with $L_{int}=1$ ab$^{-1}$ by assuming a non-zero dimension-eight operator at a time for $pp\to \nu\bar \nu\gamma$ process.  \label{tab7}}\begin{ruledtabular}
\begin{tabular}{lccc}
Couplings (TeV$^{-4}$) &$\delta_{sys}=0$&$\delta_{sys}=5\%$&$\delta_{sys}=10\%$ \\ \hline 
$C_{\widetilde B W}/\Lambda^4$  & [-0.11;0.11] & [-0.31;0.31] & [-0.44;0.44]\\\
$C_{B B}/\Lambda^4$  &   [-0.092;0.092]&  [-0.26;0.26] & [-0.37;0.37] \\
\end{tabular}
\end{ruledtabular}
\end{table}

 \begin{table}
\caption{  Obtained limits on $C_{\widetilde B W}/\Lambda^4$ and $C_{ B B}/\Lambda^4$ at 95\% C.L. with $L_{int}=3$ ab$^{-1}$ by assuming a non-zero dimension-eight operator at a time for $pp\to \nu\bar \nu\gamma$ process.  \label{tab8}}\begin{ruledtabular}
\begin{tabular}{lccc}
Couplings (TeV$^{-4}$) &$\delta_{sys}=0$&$\delta_{sys}=5\%$&$\delta_{sys}=10\%$ \\ \hline 
$C_{\widetilde B W}/\Lambda^4$  & [-0.084;0.084] & [-0.31;0.31] & [-0.44;0.44]\\
$C_{B B}/\Lambda^4$  &   [-0.072;0.072]&  [-0.26;0.26] & [-0.37;0.37] \\
\end{tabular}
\end{ruledtabular}
\end{table}


\end{document}